# Design of a Trigger Data Serializer ASIC for the Upgrade of the ATLAS Forward Muon Spectrometer

Jinhong Wang, Liang Guan, J. W. Chapman, Bing Zhou, and Junjie Zhu.

*Abstract*—The small-strip Thin Gap Chamber (sTGC) will be used for both triggering and precision tracking purposes in the upgrade of the ATLAS forward muon spectrometer. Both sTGC pad and strip detectors are readout by a Trigger Data Serializer (TDS) ASIC in the trigger path. This ASIC has two operation modes to prepare trigger data from pad and strip detectors respectively. The pad mode (pad-TDS) collects the firing status for up to 104 pads from one detector layer and transmits the data at 4.8 Gbps to the pad trigger extractor every 25 ns. The pad trigger extractor collects pad-TDS data from eight detector layers and defines a region of interest along the path of a muon candidate. In the strip mode (strip-TDS), the deposited charges from up to 128 strips are buffered, time-stamped, and a trigger matching procedure is performed to read out strips underneath the region of interest. The strip-TDS output is also transmitted at 4.8 Gbps to the following FPGA processing circuits. Details about the ASIC design and test results are presented in this paper.

## I. INTRODUCTION

ATLAS plans several major improvements in conjunction with the upgrades of the Large Hardon Collider (LHC) over the next decade. These improvements will make the ATLAS detector benefit from the increased instantaneous and integrated luminosity for new discoveries. The innermost station of the ATLAS forward muon spectrometer will be replaced by a so-called "New Small Wheel" (NSW) detector in 2019 [1]. The NSW is composed of eight layers of Micromegas chambers (MM) and eight layers of small-strip Thin Gap Chambers (sTGC) arranged in multilayers of two quadruplets, for a total active surface of more than 2500 m². Both detectors will provide trigger and tracking primitives to the muon trigger and readout system. The sTGC will be used as the primary trigger device, and the work described in this paper pertains to the sTGC front-end electronics.

The sTGC is a gaseous, multi-wire proportional drift chamber. A block diagram of the sTGC detector is shown in Fig. 1. The basic structure consists of a grid of gold-plated tungsten anode wires sandwiched between two resistive cathode planes. The wire pitch is 1.8 mm and the

Manuscript received on June 18th, 2017. This work is supported by the Department Of Energy under contracts DESC0008062 and DE-AC02-98CH10886.

The authors are with Department of Physics, University of Michigan, 450 Church Street, Ann Arbor, Michigan, 48109, US. (e-mail: jinhong@umich.edu).

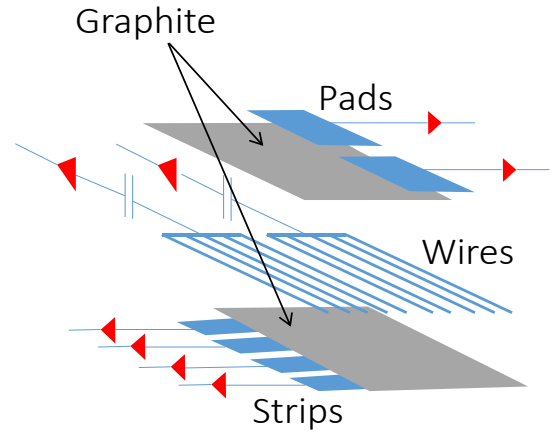

Fig. 1. Structure of the sTGC detector.

wire-to-cathode distance is 1.4 mm. One cathode is covered with ~8 cm × 8cm pads, while the other cathode is covered by strips with 3.2 mm in pitch and from 0.5 m to 2 m length. Due to the short wire-to-cathode distance, the duration of the muon hit is often less than 25 ns. Signals from five wires are grouped together to provide the hit position in the azimuthal direction. Only pads and strips are used in the trigger path.

A block diagram of the sTGC trigger electronics is shown in Fig. 2. Both sTGC pad and strip detector signals are processed by a 64-channel Amplifier Shaper Discriminator (ASD) ASIC [2]. The digitized ASD output are sent to a Trigger Data Serializer (TDS) ASIC. The TDS ASIC has two operation modes to handle pad and strip detectors, denoted here as pad-TDS and strip-TDS respectively. The pad-TDS checks the presence of pad signals, prepares and sends the pad trigger data as well as the LHC Bunching Crossing Identification number (BCID), to the pad-trigger extractor board on the rim of the NSW detector at a rate of 4.8 Gbps. The pad-trigger extractor board collects pad-TDS data from eight sTGC layers and determines a region of interest (ROI) for the candidate muon track in the form of a BCID and position coordinate (band-phi ID). The ROI is then encoded and sent to the strip-TDS at a rate of 640 Mbps. The strip-TDS decodes the deposited charge on the strips and stores the charge information together with a time tag in buffers. Once it receives the ROI from the pad-trigger extractor board, it translates the ROI to a band of strips using a pre-assigned lookup table. Pad-strip matching is performed and charges for matched strips are packed and serialized to the signal router board on the rim of the NSW detector at 4.8 Gbps

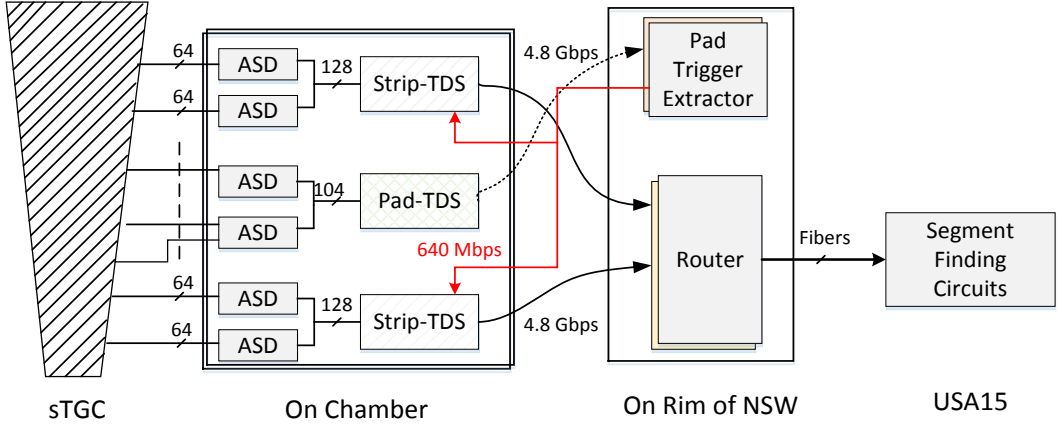

Fig. 2. Simplified block diagram of the sTGC trigger logic in NSW (only two strip TDS and one pad TDS are shown).

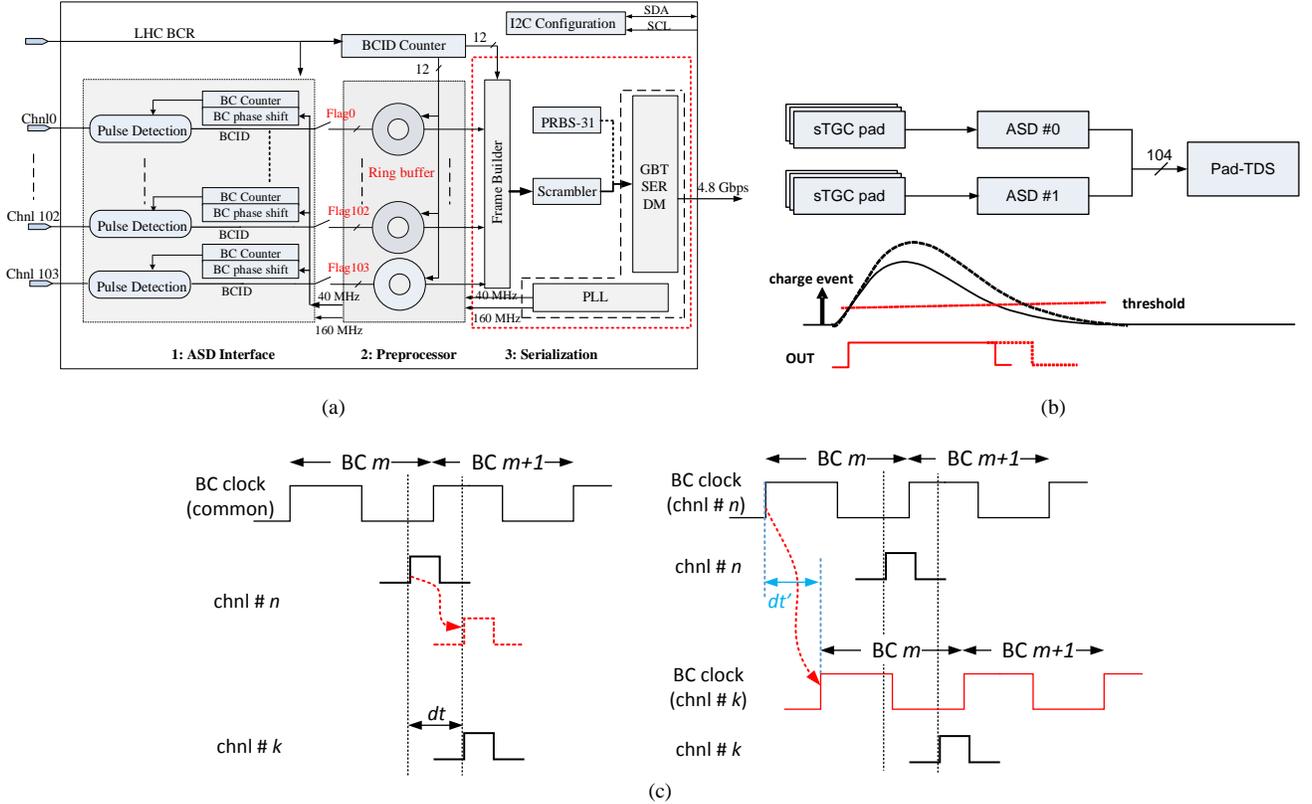

(a)

(b)

(c)

Fig. 3. The pad-TDS ASIC: (a) the block diagram; (b) ASD inputs to the pad-TDS; (c) illustration of two delay compensation schemes: inserting delay elements (left) and shifting the timing clock phases (right).

[3]. The router board collects strip trigger data from up to twelve strip-TDS ASICs, removes NULL packets and transmits the data via optical fibers to the segment-finding circuits in the ATLAS underground counting room (USA15).

Each TDS handles 128 channels in the strip mode and 104 channels in the pad mode. In addition to the high channel density, the ASIC is also required to have a power consumption less than 1 W and a global latency less than 100 ns. The ASIC also needs to work in a harsh radiation environment for about 20 years. These requirements all impose challenges in the TDS design.

The paper is organized as follows: we describe the design of pad-TDS and strip-TDS in Section II and III, respectively. The prototype and test results are presented in Section IV. There are further discussions of the results in Section V, followed by a conclusion in Section VI.

## II. DESIGN OF THE PAD-TDS

A block diagram of the pad-TDS is shown in Fig. 3(a). The design is divided into three parts: ASD interface, Preprocessor and Serialization. These three parts are described in detail in the following subsections.

### A. ASD Interface of the pad-TDS

The ASD interface of the pad-TDS captures the pad firing status from up to 104 sTGC pads (maximum number of pads in an sTGC quadruplet) and adds a BCID time tag for each channel. Since one ASD can only process up to 64 pads, two ASDs are used, as shown in Fig. 3(b). The pad hit signal is amplified, shaped and discriminated inside the ASD, and a Time-Over-Threshold (TOT) pulse is sent to the TDS. The BCID time tag is assigned upon the arrival time of the leading edge of the TOT pulse.

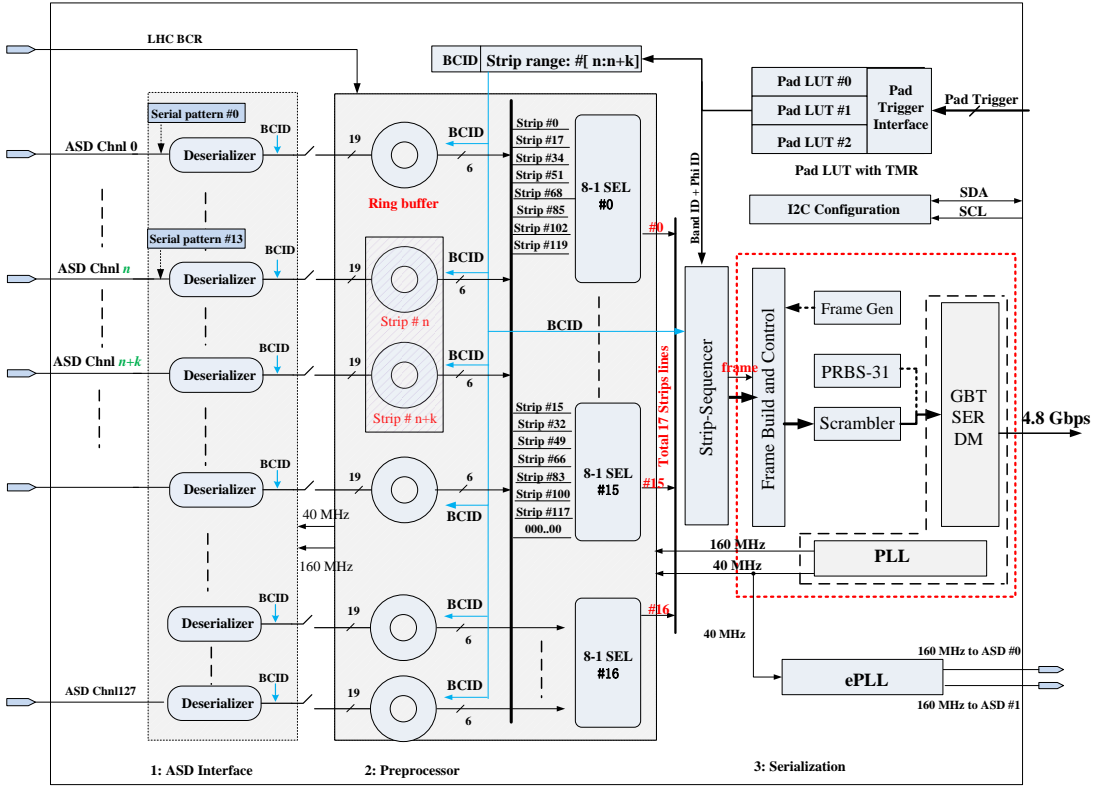

Fig. 4. Simplified block diagram of the strip-TDS.

Since one TDS handles up to 104 pads inside an area of about 2 m², the routing length from a pad to the pad-TDS can be different by up to 3 meters. This routing length difference could result in different BCID time tags for pad signals with a common arrival time. This difference needs to be compensated for each channel. The time compensation precision is required to be 3.125 ns for a range up to 25 ns. Conventionally, additional delay is inserted into early arrival channels, as shown in the left diagram of Fig. 3(c). However, the usage of either pure delay cells or feedback controlled delay circuits (e.g. with Delay Lock Loops) is excessive from the point view of the logic resource usage, system complexity and power consumption for 104 channels. Alternatively, we compensated the delay by adjusting the phase of timing clocks (BC clock), as shown in the right part of Fig. 3(c). Individual timing clock for each channel is utilized and the phase shift is achieved by BC clock regeneration through shift registers running at dual edges of a

160 MHz clock [5]. This scheme costs only a few flip flops and is fully synthesizable. Its stability is inherent to that of the 160 MHz clock.

### B. Preprocessor of the pad-TDS

The Preprocessor circuit buffers the time-tagged pad signals in a 2-depth ring buffer for each channel, and compares the current BCID to those of the buffered signals to flag the YES/NO firing status of every channel at each BCID. A YES is marked if any hits with the same BCID are found, otherwise a NO is flagged.

The ring buffer is built from a two-stage shift register: whenever a new hit is written into the buffer, registers at both stages advance forward with the second buffer pushed out to free the first slot for the new hit. A timer is used to monitor the buffer status and to avoid a pad hit occupying the slot indefinitely in case there are no pad hits for a long time. A NULL signal is pushed into the ring buffer at a programmable count of the timer to kick out the older hit. BCID comparison and firing status assignment is done at the end of a BC and the referred BCID is given by a global BCID counter as shown in Fig. 3(a).

### C. Serialization of the pad-TDS

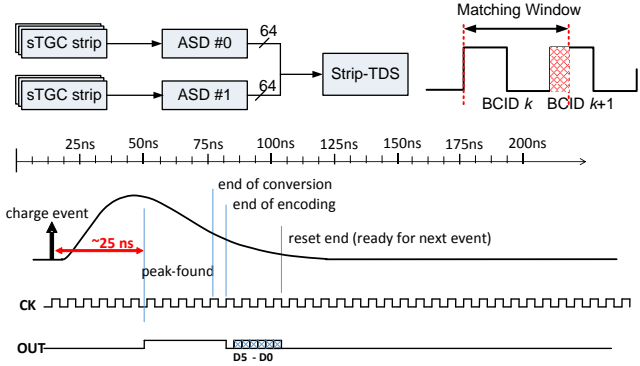

Fig. 5. ASD inputs to strip-TDS.

The firing statuses of all 104 channels are collected at each BC and the result is passed to the "Frame Builder" circuit which combines the 12-bit BCID with 104 pad firing status bits to form a 116-bit packet. These 116 bits are spilt into four consecutive frames: a 26-bit frame followed by three 30-bit frames. The four frames are scrambled following a scheme used in the IEEE Standard 802.3-2012 for 10 Gbps physical layer implementation (with a polynomial function of $1+x^{39}+x^{58}$) [6]. A 4-bit header "1010" is added to the beginning of the 26-bit scrambled frame, making it the first of the four 30-bit frames to be loaded into the serializer core (GBT SER DM) [4]. This serializer core loads inputs at 160 MHz and serializes 120 bits every 25 ns. It is a modified version of the original CERN design [7]. In addition, there is a pseudorandom binary sequence generation with every permutation of 31 bits (PRBS-31 with a polynomial function of $x^{31}+x^{28}+1$) for the test purpose of the 4.8 Gbps serializer interface.

## III. DESIGN OF THE STRIP-TDS

A block diagram of the strip-TDS is shown in Fig. 4. Similar to the pad TDS, the architecture is divided into three stages: ASD interface, Preprocessor and Serialization.

### A. ASD Interface of the strip-TDS

The ASD interface handles 128 strip outputs from two ASDs. Figure 5 illustrates the typical timing and waveform of the ASD output to the strip-TDS. The strip charge signal is first amplified and shaped. Once a peak is found, the ASD output (*OUT*) turns high and a 6-bit ADC digitizes its value. Upon completion of the digitization, *OUT* will be pulled low following the rising edge of the clock (*CK*) for half a clock cycle, and then the 6-bit charge is serialized to the strip-TDS at dual edges of *CK*. *CK* is the 160 MHz clock provided by the ePLL circuit [8] as shown in Fig. 4, leading to a line rate 320 Mbps for *OUT*. For the strip-TDS test purpose, an embedded ASD charge pattern generator for the first 14 channels is implemented.

The serial charge input is decoded by a *Deserializer* in each channel with a BCID time tag attached. The BCID is determined by the first leading edge of *OUT*, which corresponds to the peaking time. In addition to the charge and the BCID tag, a "FLAG" bit is included in case a different trigger matching time window is used. Since the variations of the charge arrival time on different strips for the same muon hit can be larger than one BC clock cycle, the trigger matching time window is chosen to be programmable to maximize the charge matching efficiency. The matching window is programmable from 25 ns to 50 ns, with a step size of 6.25 ns. For a given matching window as shown on the top right of Fig. 5, any signal arriving at the beginning of BCID $k+1$ (the dashed area) might also belong to BCID $k$ if the charge signal has a longer drift time. An "FLAG" bit is introduced: for any signal arriving in the dashed area of BCID $k+1$, its "FLAG" is set with a logic TRUE, otherwise it is set to FALSE. Finally, the 6-bit charge, BCID and the "FLAG" bit are packed as a "charge strip data unit" to be sent forward.

### B. Preprocessor of the strip-TDS

The charge strips obtained from the prior stage are buffered in the ring buffer using four shift registers (BUF0-3) for each channel, as shown in Fig. 6(a). Every time there is a trigger matching request for a given channel, the shift registers are sampled for further processing in the "BCID comparison" unit. The ring buffer is data-driven and a programmable timer monitors buffer status to avoid any events staying in the memory slot forever.

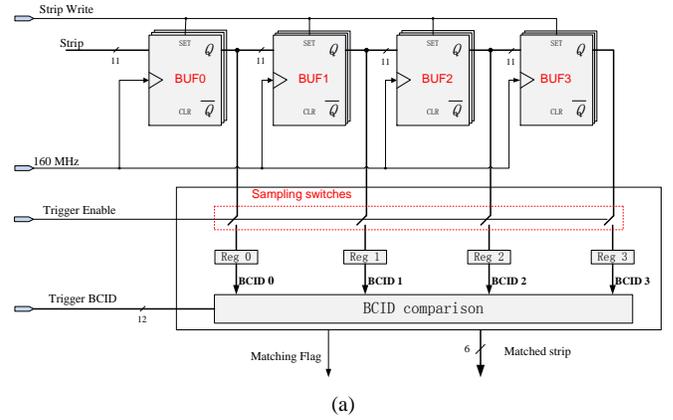

(a)

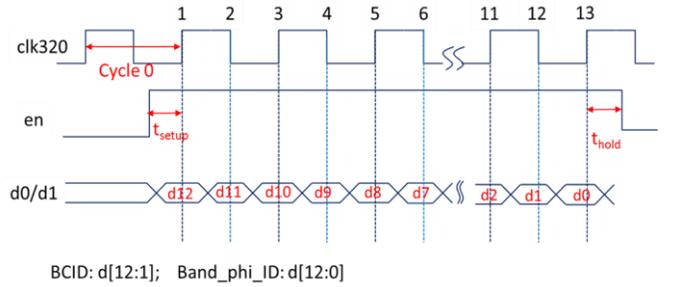

(b)

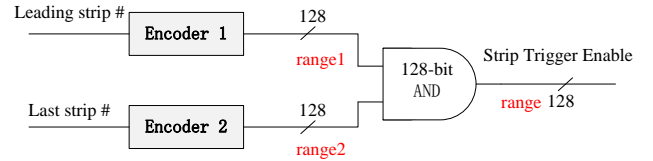

(c)

Fig. 6. (a) Ring buffer and trigger-matching unit for strip-TDS; (b) timing of pad-trigger interface; (c) Channel trigger-enable algorithm.

The trigger request comes from the pad trigger extractor board on the rim of the NSW as shown in Fig. 2. The trigger request to each TDS is made via four lines: a 320MHz clock (*clk320*), an enable (*en*) signal, and two data lines at 640 Mbps for the trigger BCID and band-phi ID (ROI position coordinate) respectively (*d0/d1*), as shown in Fig. 6(b). The pad trigger interface circuit inside the strip-TDS uses the 320 MHz clock to decode the trigger BCID and band-phi ID information. The band-phi ID corresponds to the coordinates of the extracted muon track, which is then translated into a band of strip channels via the pre-assigned lookup tables. Strips in this band are selected and matched charges on these strips are readout. The strip selection is done by encoding the channel number of the first (leading strip) and last strip in the band with a 128-bit thermal code (*vt*[0:127]) and inverse thermal code (*vt_inv*[0:127]), respectively. A logic "AND" of the two vectors results in a 128-bit vector, corresponding to the trigger selection of strip channel 0-127 from the least significant bit (LSB). One example with a band of strip numbers from 10 to 26 is shown in Fig. 6(c), in which "10" is encoded as a 128-bit thermal code: *vt*[0:9]="00…0", *vt*[10:127]="11…1"; "26" is encoded as *vt_inv*[0:26]="11…1", *vt_inv*[27:127]="00…0". The 128-bit vector after logic "AND" has "1"s only on bit positions 10-26, thus strip channel 10-26 are enabled for trigger matching in this case. For all channels within the band enabled for trigger matching, BCID comparison is performed with the FLAG bits for the sampled buffers in the "BCID comparison" unit. When the BCID timing is matched for a given channel, a valid matching signal is generated and the corresponding charge is picked up for further processing.

A sTGC pad covers about 13 strips. In addition, two neighboring strips at the two edges (top and bottom) have to be considered since a muon could hit the pad boundaries. As a result, a maximum of 17 consecutive strips will be enabled in a trigger matching. An efficient way to multiplex 17 out of 128 strips for any trigger band of strips is needed. To achieve this, all 128 channels are divided into seventeen groups with eight channels in each group using seventeen 8-1 selectors. For example, the first 8-1 selector connects channels 0/17/34/51/68/85/102/119, and the second 8-1 selector connects channels 1/18/35/52/69/86/103/120, as shown in Fig. 4. This arrangement guarantees that all 17 strips in a band will not be routed to the same multiplexer. However, the multiplexer outputs in this configuration might not preserve the original sequence as in the trigger band, e.g. the multiplexed outputs from SEL #0-16 for a band with strip #18-34 is 34, 18, 19, … 33, in which strip #34 is moved to the front. The strip sequence problem is resolved by the "*Strip-Sequencer*" circuit as the leading strip is known from the trigger.

### C. Serialization of the strip-TDS

The band of strips after the trigger matching will be passed to the Serialization circuit. The "Frame Build and Control" unit collects charges from 17 strips as well as other time and ROI information and builds a 120-bit packet, which will be sent out to the Router board in 25 ns at 4.8 Gbps. The serial protocol is shown in Fig. 7(a), in which the data are arranged in four consecutive 30-bit frames. NULL frames are inserted when there are no trigger matching requests from the pad trigger extractor. Each 30-bit frame starts with a 4-bit header. Different headers for data and NULL are used for quick data frame switching on the sTGC Router, where NULL frames from up to 12 TDS links are suppressed and data frames are merged to be sent out via four optical links.

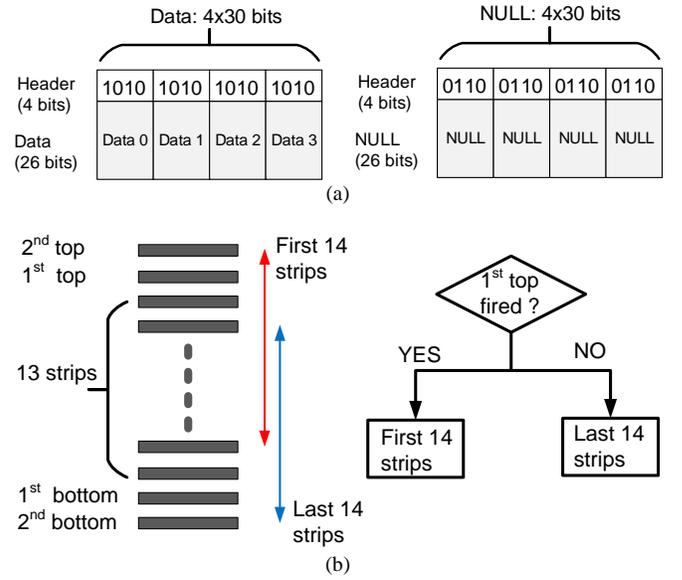

Fig. 7. (a) strip-TDS output serial protocol; (b) selection 14 strips out of the 17 strips in a trigger.

The time and ROI information that needs to be transmitted are 13 bits of band-phi ID and 6 bits of trigger BCID from the LSB, which leaves a maximum of 85 bits for the strip charges. Since 6 bits are used for each strip charge, the number of strips to be readout in a band is reduced to 14. This is fine since in reality the size of a muon cluster is around 4-5 strips, and it is not necessary to send out the charges for all 17 strips. The reduction is performed upon the firing status of the top first neighboring strip to select either the first or the last 14 strips of the 17 strips, as shown in Fig. 7(b). An additional bit is included to indicate which 14 strips are selected. The total number of bits used is then 14 strip × 6 bit/strip + 1 =85.

The content of a 30-bit frame except the header is scrambled [6] before loading into the serializer core [4]. A fake TDS-Router protocol generator is implemented to build the TDS Router link and to test the complete serial link from TDS to the USA15 through the Router. A PRBS-31 generator is also provided for the serial link evaluation.

### IV. TDS PROTOTYPES

The TDS design started in 2013 with the serializer-only chip prototype [4], which was a crucial part of the pad-TDS and strip-TDS. The prototype proved to be successful in early 2014. The first prototype of TDS (TDSV1) was submitted in August 2014 and also met design specifications. A second prototype

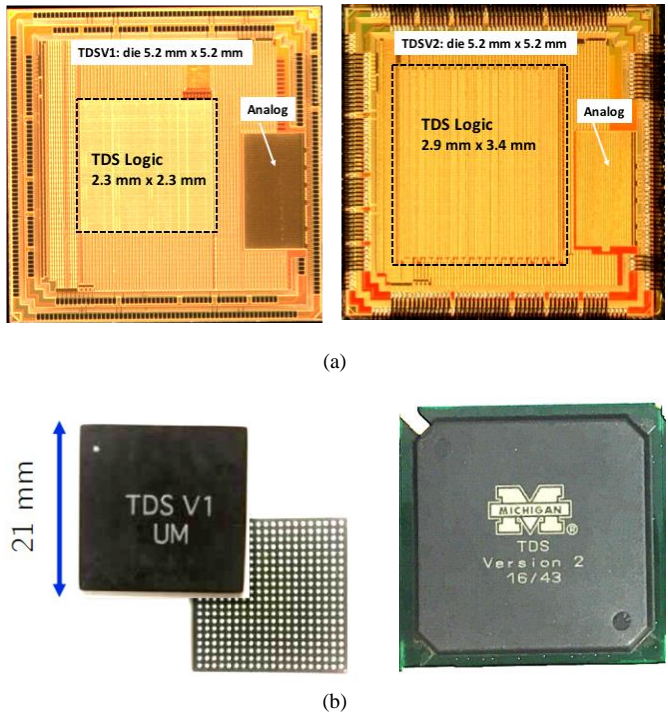

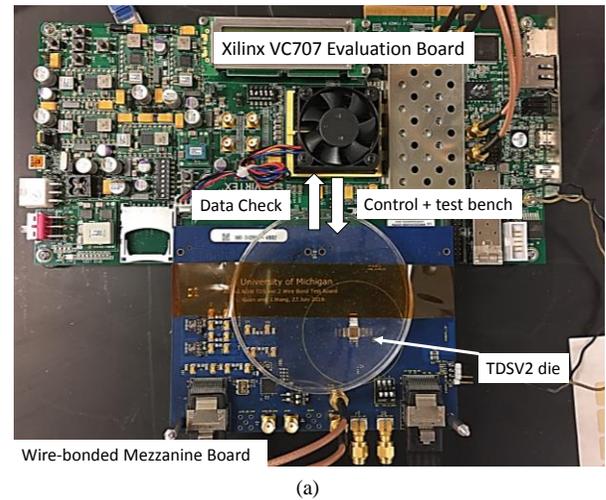

(a)

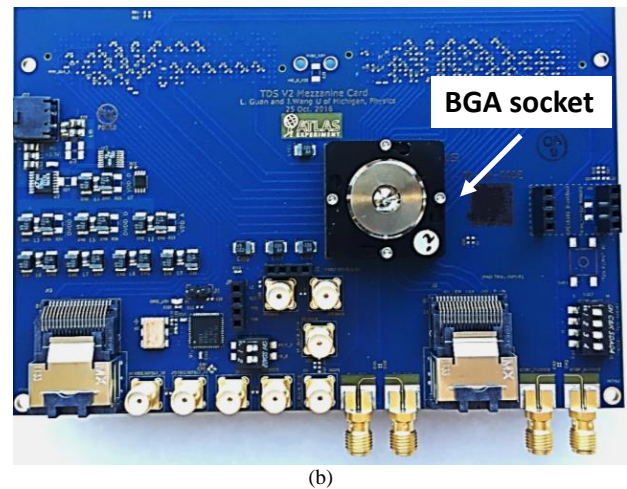

(b)

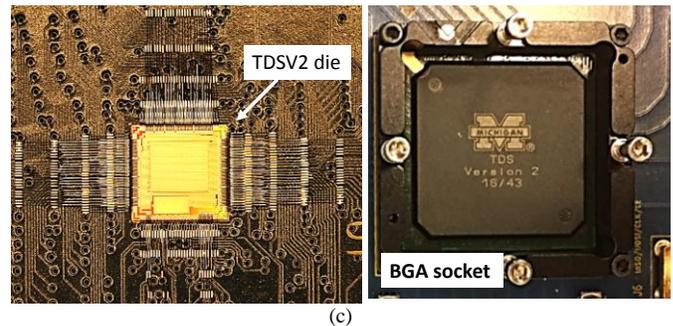

(c)

Fig. 9. (a) The TDSV2 wire-bond mezzanine board with the Xilinx VC707 board; (b) Photograph of the TDSV2 BGA socket board; (c) TDSV2 die directly bonded to a PCB (left) and in a BGA socket (right).

Fig. 8. (a) Pictures of TDS dies for prototype version 1 (left) and version 2 (right); (b) Packages for the TDS version 1 and version 2.

(TDSV2) was submitted in May 2015 to accommodate more specification changes. Pictures of the two prototype dies are shown in Fig. 8(a), in which the "TDS Logic" part includes all logic processing for the pad and strip modes. The area of the "TDS Logic" is almost doubled in TDSV2 with respect to that in TDSV1. A comparison of the logic resource utilization for TDSV2 and TDSV1 after the synthesis and their ratio is shown in Table 1, in which about 1.8 times more cells are used in TDSV2. The ratio is consistent with the "TDS Logic" area which increased from 5.29 mm² to 9.86 mm². The "Analog" part includes the serializer core, ePLL and the pad-trigger interface. The first two parts are the same in both prototypes while the pad-trigger interface was modified in TDSV2 for a new protocol as defined in Fig. 6(b). The area of the "Analog" part and also the size of both dies remain the same (5.2 mm × 5.2 mm) for both prototypes.

The TDS is packaged in a 400-pin Ball Grid Array (BGA) package as shown in Fig. 8(b). In total, 191 TDSV2 dies were produced and 161 of them were packaged in BGA for the TDSV2 prototype.

## V. TDS PERFORMANCE EVALUATION

### A. Test Setup

The TDS utilizes a wire-bond 400 pin BGA package. There are several stages in the design of the BGA package including the design of a substrate accommodating the chip die. It takes several months to complete the whole procedure. The TDS performance was first evaluated by direct wire-bonding to a test board, as shown in Fig. 9(a). The wire-bonded test board was designed as a mezzanine card to a Xilinx VC707 evaluation board. The Virtex-7 FPGA on the VC707 board provides all test inputs, configuration controls and also receives the 4.8 Gbps output from the TDS for data checking. The connections are done via the pair of High-Pin-Count (HPC) connectors on the VC707 board. A photograph of the wire-bonded TDSV2 is shown on the left plot of Fig. 9(c).

For the packaged TDS, a mezzanine card with a BGA socket was also designed, as shown in Fig. 9(b). The BGA socket replaces the wire-bonded die on the wire-bonded mezzanine and other parts remains the same. A picture of the BGA socket with a packaged TDS ASIC inside is shown on the right plot of

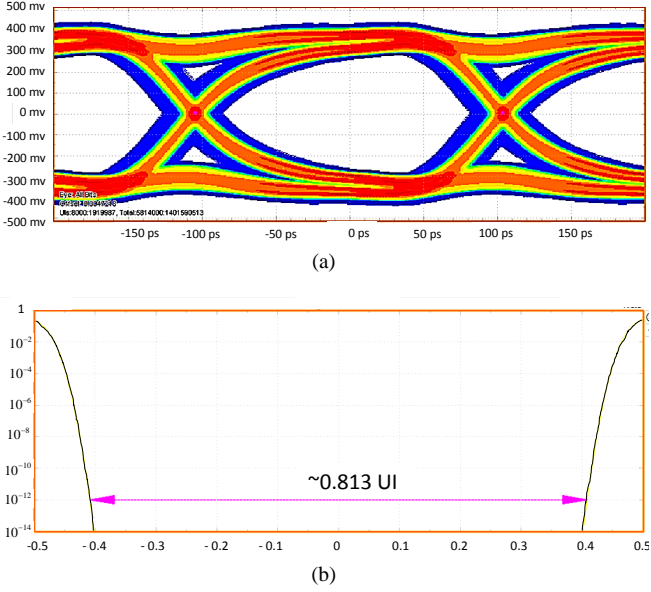

Fig. 10. Performance of the serializer with PRBS-31 pattern, where (a) is the eye diagram; (b) is the bathtub curve.

TABLE I
LOGIC RESOURCE UTILIZATION

| Type | TDSV1 | TDSV2 | Ratio (TDSV2/TDSV1) |
|------|-------|-------|---------------------|
| Sequential | 47451 | 76772 | 1.62 |
| Inverter | 9331 | 23322 | 2.45 |
| Buffer | 424 | 399 | 0.94 |
| Logic[*] | 104841 | 193973 | 1.85 |
| Total | 162047 | 294466 | 1.82 |

[*]Logic: combinational logic cells in standard library, such as "AND", "OR" gates.

TABLE II
TEST COVERAGE OF EMBEDDED TEST

| Test Mode | TDS-Router Training Frame | Global-Test | Bypass Trigger |
|-----------|---------------------------|-------------|----------------|
| Serial Pattern | × | √ | × |
| Deserializer | × | √ | √ |
| Ring buffer | × | √ | × |
| Trigger Matching | × | √ | × |
| Pad-trigger interface | × | × | × |
| 8-1 selector | × | √ | √ |
| Sequencer | × | √ | √ |
| Serializer | √ | √ | √ |

√ : covered; × : not covered

Fig. 9(c). All test results presented in this paper are about the TDSV2 with the BGA mezzanine card.

## B. Performance of the Serializer

The performance of the serializer core (GBT SER DM) has been evaluated in the serializer-only prototype [4]. However, the noisy mixed-signal environment of TDS may introduce extra noise and degrade the performance of the serializer core. To minimize induced noise from the digital circuits, the silicon substrate of the serializer core is isolated from the "TDS Logic" and separate power and ground grids are used for the "Analog" part in Fig. 8(a).

All tests in [4] were repeated and yielded the eye diagram and the bathtub plot of the serializer output with embedded PRBS-31 pattern in Fig. 10. Further jitter analysis shows a total jitter of about 39 ps, which is slightly better than the 49.7 ps reported in [4]. These results indicate that the performance of the serializer core has not been impacted by integration into the TDS ASIC. These tests are evaluated with a Tektronix 12.5 GHz bandwidth 50 GS/s oscilloscope (*DSA71254B*).

## C. Performance of the strip-TDS

The strip-TDS utilizes a trigger scheme to select strips underneath a band determined by the pad trigger extractor board, which makes the use of strip mode data only available upon a trigger request. The complicated trigger matching process makes it difficult to probe individual functional blocks for diagnosis along the signal chain (as shown in Fig. 4). Several embedded test functions are added in the TDS design. These test functions address critical function blocks while simplifying or bypassing other remaining blocks for direct output. A summary of the coverage of three major test functions is shown in Table II.

The "TDS-Router Training Frame" corresponds to the "*Frame Gen*" in Fig. 4. It generates fake test frames according to the TDS-Router protocol shown in Fig. 7(a) and can be used for establishing the link between the TDS and the Router. This mode only covers the serialization stage of the strip-TDS. The "Global-Test" mode utilizes internal serial pattern generator to create the ASD output that goes through the whole signal flow except the "Pad-trigger interface". In this mode, internal triggers are used. "Bypass Trigger" mode works with external ASD inputs which bypasses the channel ring buffer and trigger units. Accompanied with the single channel enable, the "Bypass Trigger" mode can probe the individual channel inputs without any trigger requests. Combinations of these test functions can test all functions of the strip-TDS except the pad trigger interface.

To fully characterize the performance of the strip-TDS, we simulate inside a Xilinx Vertex 7 FPGA, the ASD inputs and the pad-trigger generator for the strip TDS. In the test, the charge of each of the 128 channels is assigned as the least significant six bits of its channel number, e.g., channel 15 has a charge of 15 while channel 67 has a charge of 3. These inputs are induced at a fixed BCID (e.g. BCID = 36) to simplify the analysis, and four consecutive triggers separated by 25 ns each (one BC cycle) are given around BCID 36 in this example. For each trigger, a corresponding range of strips is evaluated and the lookup table in Fig. 4 is configured such that all 128 channels are covered. Typical results are shown in Fig. 11. For a matching window of 25 ns where an exact BCID comparison is performed, only channels with BCID 36 have matches as shown in Fig. 11(a); whereas if we extend the matching window to 50 ns, strips with BCID 36 have the BCID "FLAG" bit tagged as TRUE, thus satisfy the trigger matching requirement with trigger BCID 35, as shown in Fig. 11(b).

## D. Performance of the pad-TDS

The performance of the pad-TDS is evaluated by simulating the pad inputs in VC707 and decoding the serial outputs to check the corresponding presence. For simplicity, hits to all channels are generated at the same time. Additional delay is introduced as the pad signal traverses through the FPGA output buffers. The delays introduce up to 3 ns variances between channels. In the test, the delay is used to tune the phase of the test bench clock so that some of the channels falls to one BCID (0x017) while the others in the next one (0x018), as shown in Fig. 12(a). Delay compensation is then added for all channels with BCID 0x018 to cancel the extra delay in its path. As a result, all inputs converge into one BCID (0x017), as shown in Fig. 12(b). In total 40 pad channels are shown in Fig. 12 as an example (channel 73-103) with the help of a virtual oscilloscope in the FPGA. For each channel at a given BCID, the channel flags "1" if there is a pad hit presenting in that BCID, otherwise it stays at "0". Detailed characterization of the pad delay compensation is reported in [5].

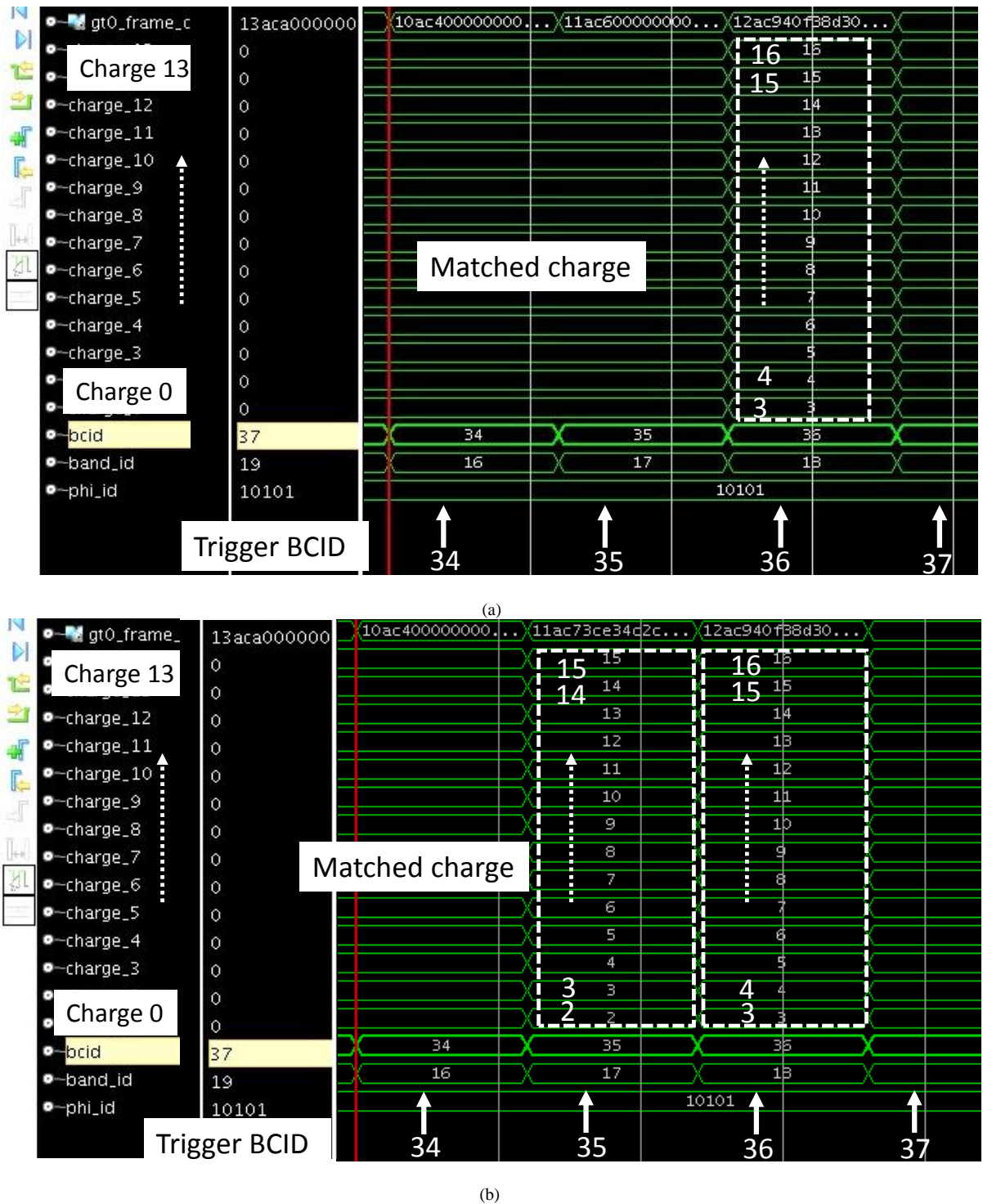

(a)

(b)

Fig. 11. Typical consecutive trigger tests of Strip-TDS where (a) shows results with a 25 ns matching window and (b) with 50 ns matching window.

### E. Power Consumption

The TDS uses a 1.5 volt supply for both the logic part and the

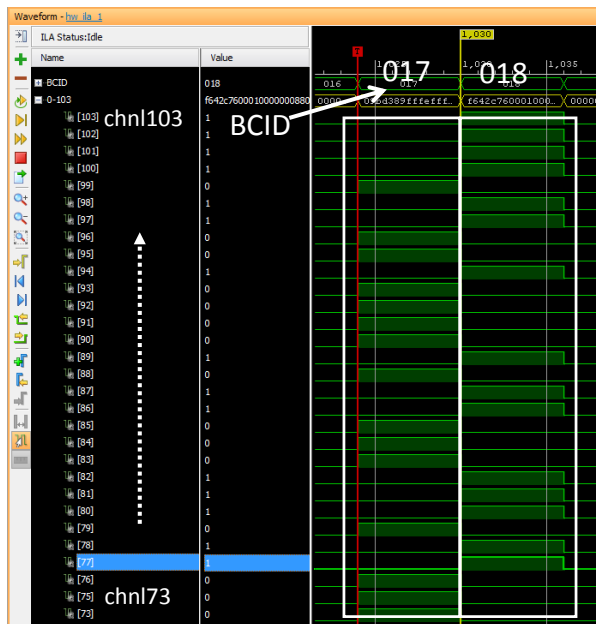

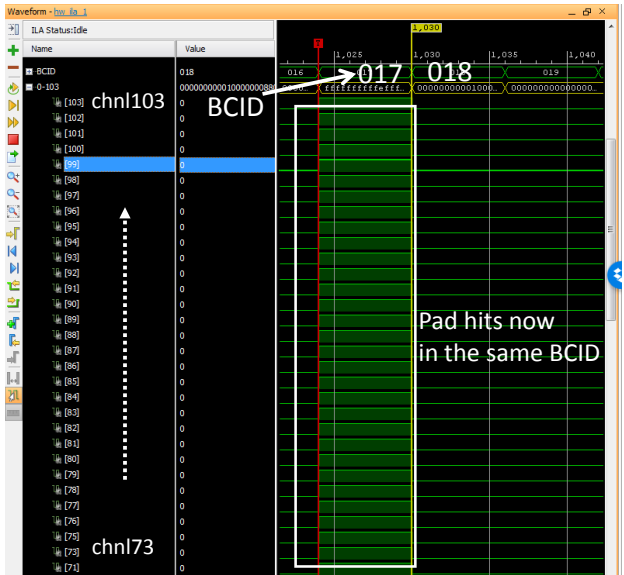

Fig. 12. Typical result of pad TDS, (a) before channel delay compensation; (b) after delay compensation.

| Individual Stage | Latency | Note |
|---|---|---|
| Clock crossing | 6.25 ns (max.) | From clk320 to clk160 |
| Find strips in ROI | 12.5 ns | Look through LUT |
| Trigger Matching | 18.75 ns | Ring buffer sampled |
| Strip Sequencer | 12.5 ns | Resolve the sequence |
| Build Frame | 12.5 ns | Prepare serializer frames |
| Scrambler | 6.25 ns | DC balance |
| Serializer core | ~6 ns [4] | First bit in to first bit out |

TABLE IV
ALLOCATION OF THE TOTAL LATENCY OF THE PAD-TDS

| Individual Stage | Latency | Note |
|---|---|---|
| Collect pad status | 6.25 ns | For all 104 channels |
| Build Frame | 12.5 ns | Prepare serializer frames |
| Scrambler | 6.25 ns | DC balance |
| Serializer core | ~6 ns [4] | First bit in to first bit out |

## VI. DISCUSSION

### A. Latency

Both pad and strip modes of the TDS work with a global clock of 160 MHz (clk160), as shown in Fig. 3 and Fig. 4. Different stages along the signal path from the input to the output are put in pipeline of clk160, and the total latency is allocated into each stage. Design optimization is evaluated to meet the total latency requirements.

The latency of the strip-TDS starts from the time of any trigger request at the pad trigger interface to the time that the first bit of data frame exits the serializer core. The latency breakdown along the signal path is listed in Table III. The total latency is found to be 75 ns, satisfying the requirement of 100 ns.

The latency of the pad-TDS starts from the end of a BC to the first bit of a pad frame exiting the serializer core. The total latency is itemized in Table IV and the overall latency is found to be 31 ns.

### B. Radiation Tolerance

The Total Ionizing Dose (TID) for the NSW sTGC trigger front-end boards is about 300 kRad with a safety factor of 6 [9]. This radiation exposure poses no problem for the 130 nm CMOS process that was used [10]. For protection of Single Event Upsets (SEU), Triple Modular Redundancy (TMR) is applied to both the analog part (the serializer core and ePLL) [4] and the TDS logic. In the TDS logic, full TMR is implemented with three voters and also triple clock trees.

### C. Preparation for Production Test

TDSV2 is proven to be complete and meets all design specifications. The 161 packaged TDS ASICs were tested with the BGA socket mezzanine card as shown in Fig. 9(b). Automatic test procedure are developed, in which the FPGA on VC707 takes control of the whole test procedure. Three test stations were running in parallel to test all 161 chips, and 158 chips passed all functional tests. The yield is found to be about 98%. More statics will be collected in the final production where 6,000 chips will be produced and tested.

serializer. The power consumption is evaluated by putting a 0.1 Ohm (1%) resistor in serial with the power supply and monitoring the voltage drop. The voltage drop is 57.8 mV for the strip mode and 58.5 mV for the pad mode, which corresponds to a current of 578 mA and 585 mA respectively. The corresponding power consumption of the strip-TDS and the pad-TDS is therefore 0.867 W and 0.878 W, respectively. The mode of the TDS, pad or strip, is selected via a configuration pin. While one mode is active, the other will be reset to save power. The values measured represent the total power consumption in each mode.

## VII. Conclusion

The TDS is a mixed signal ASIC with two operation modes for the sTGC pad and strip detectors at ATLAS. Detailed signal processing inside the chip was introduced for both modes and preliminary test results for the TDSV2 ASIC were reported.

## Acknowledgment

The authors would like to thank P. Moreira, R. Francisco, and F. Tavernier from CERN; De Geronimo Gianluigi, Daniel Levin from the University of Michigan; T. Liu and D. Gong from Southern Methodist University; Lorne Levinson from Weizmann Institute of Science for their help in this work.